%% file: main.tex
\documentclass[final,pdf,3p,twocolumn,times,11pt]{elsarticle}

\pdfoutput=1

\usepackage{amsmath, amssymb}

\usepackage{xspace}
\usepackage{tensor}
\usepackage{color}

\include{notation}

\journal{Physics Letters B}

\begin{document}

\newif\ifdraft
\drafttrue
\def\draftnote#1{\ifdraft{\it #1}\fi}
\newcommand{\tsc}[1]{\textsc{#1}}
\newcommand{\Ar}{\tsc{Ariadne}}
\newcommand{\Vc}{\tsc{Vincia}}
\newcommand{\Pythia}{\tsc{Pythia}}
\newcommand{\Fastjet}{\tsc{FastJet}}
\newcommand{\Qemit}{\ensuremath{Q_\emit}}
\newcommand{\Qstart}{\ensuremath{Q_\start}}
\newcommand{\Qstop}{\ensuremath{Q_\text{stop}}}
\newcommand{\antc}{a_c}
\newcommand{\ordant}{{\bar a}}
\newcommand{\Tr}{\mathop{\rm Tr}\nolimits}
\newcommand{\eqnref}[1]{eq.~\eqref{#1}}
\newcommand{\figref}[1]{fig.~\ref{#1}}
\renewcommand{\pT}{p_{\rm T}}
\newcommand{\kT}{k_{\rm T}}

\begin{frontmatter}



\title{Antenna Showers with Hadronic Initial States}


\author[saclay]{M.~Ritzmann}
\author[saclay]{D.~A.~Kosower}
\author[cern]{P.~Skands}

\address[saclay]{Institut de Physique Th\'eorique, CEA Saclay, 
F--91191 Gif--sur--Yvette cedex, France}
\address[cern]{CERN TH, Geneva 23, CH-1211, Switzerland}

\begin{abstract}
We present an antenna shower formalism including
contributions from initial-state partons and corresponding backwards
evolution.  We give a set of phase-space maps and antenna functions for
massless partons which define a complete shower formalism suitable for
computing observables with hadronic initial states. 
We focus on the initial-state components: 
initial--initial and initial--final antenna configurations. The  
formalism includes comprehensive possibilities for  uncertainty
estimates.  We report on some preliminary results obtained 
with an implementation in the \Vc{} antenna-shower framework.
\end{abstract}


\end{frontmatter}


\section{Introduction}
\label{sec:intro}

Parton-shower algorithms offer a universal and fully exclusive 
perturbative resummation framework for high-energy processes. In the
context of Monte Carlo event generators~\cite{Buckley:2011ms}, 
they also provide the  perturbative input for hadronization models. As such, 
they are complementary to  
more inclusive techniques, such as fixed-order calculations 
(limited to small numbers of hard and well-separated partons)
and more inclusive resummation approaches (limited to a fixed set of
observables).

Sj\"ostrand derived the first consistent parton-shower algorithm~\cite{Sjostrand:1985xi} for so-called ``backwards
evolution'' of initial-state partons a quarter-century ago.
The central point is that an initial-state 
parton defined at a high factorization scale, $Q_F$, can be evolved
``backwards'', towards earlier times, to find the parton from which
it originated at some low scale, $Q_0\sim 1\,$GeV. 
During this evolution, which is governed by the 
Altarelli-Parisi splitting kernels~\cite{Altarelli:1977zs}
supplemented by PDF ratios (a point which is crucial to the
backwards-evolution formalism), initial-state radiation is emitted, 
which in turn gives rise to its own final-state radiation, and the
character of the evolving parton changes, migrating towards
successively higher $x$ values and towards the more valence-dominated
flavor content at low $Q$.

As an alternative to Altarelli-Parisi evolution, Gustafson and
Pettersson proposed 
a final-state algorithm based on QCD dipoles~\cite{Gustafson:1987rq},
which has been implemented in 
\Ar{}~\cite{Lonnblad:1992tz}. There, however, 
initial-state radiation does not rely on backwards evolution. 
Instead, it is treated essentially as final-state radiation off
dipoles stretched between the hard process and the beam remnants, 
and thus depends on the non-perturbative makeup of 
the remnants.
Winter and Krauss took a first step towards 
combining the dipole formalism with backwards
evolution (and thus also eliminating the dependence on the remnants) 
in ref.~\cite{Winter:2007ye}.  
Our construction differs in the antenna functions, evolution variables,
and recoil strategy.  In particular, it differs in the treatment of
collinear singularities in initial--final antenn\ae{}.  We have checked
that our antenn\ae{} properly reproduce all QCD singularities.

A complementary approach which merges the Lund dipole language with
that of fixed-order antenna
factorization~\cite{Kosower:1997zr,Kosower:2003bh,
GehrmannDeRidder:2005cm, Daleo:2006xa},
is that of 
\Vc{}~\cite{Giele:2007di,GehrmannDeRidder:2011dm,LopezVillarejo:2011ap}. 
(Note: we henceforth use the
term ``antenna'' rather than ``dipole'' to avoid ambiguities of
historical origins, see e.g., ref.~\cite{Bern:2008ef}). 
So far, however, the \Vc{} formalism has been applied only to
final-state showers. 
In this paper, we present all the ingredients necessary to construct
a consistent initial-state shower based on QCD antenn\ae{}. A further important
ingredient is comprehensive possibilities for uncertainty estimates,
in line with the framework for automated theory uncertainties proposed
in ref.~\cite{Giele:2011cb}.

\section{Antenn\ae{} and Antenna Showers}

\begin{figure}[t]
\centering
\parbox[c]{0.7cm}{\colorbox{black}{\textcolor{white}{II}}}
\parbox[c]{3cm}{\centering
\includegraphics*[scale=0.6]{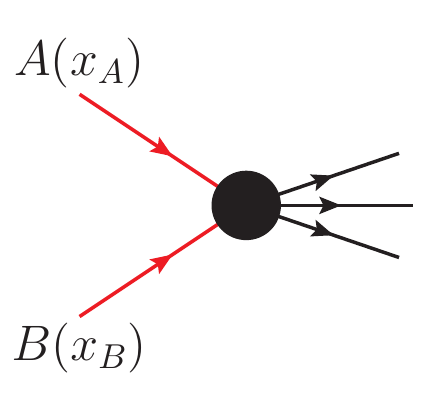}
}
\parbox[c]{0.5cm}{\centering$\to$}
\parbox[c]{3cm}{\centering
\includegraphics*[scale=0.6]{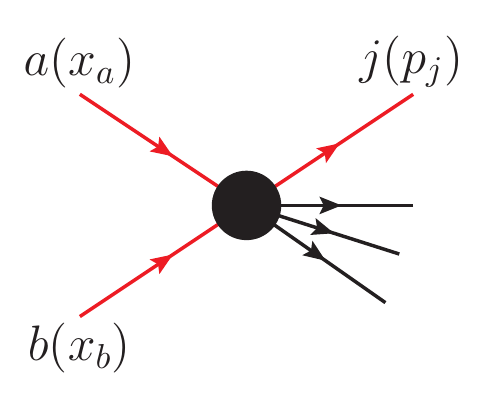}
}\\[1ex]
\parbox[c]{0.7cm}{\colorbox{black}{\textcolor{white}{IF}}}
\parbox[c]{3cm}{\centering
\includegraphics*[scale=0.6]{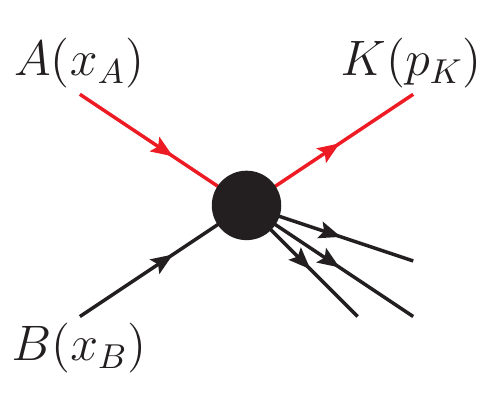}
}
\parbox[c]{0.5cm}{\centering$\to$}
\parbox[c]{3cm}{\centering
\includegraphics*[scale=0.6]{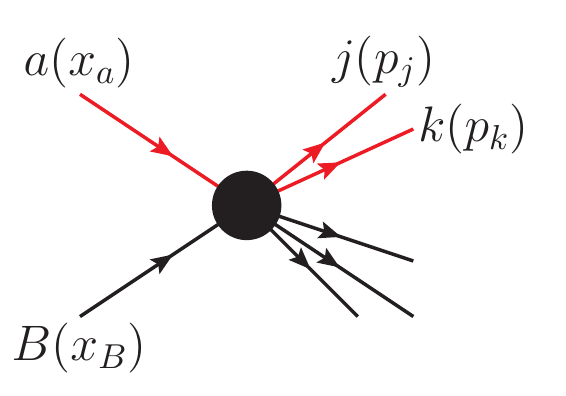}
}
\caption{Illustration of initial--initial and initial-final branchings: 
$AB \to ajb$ and $AK\to ajk$, respectively. For the II case, 
the recoil of the hard system is illustrated by the change in
  orientation of the three outgoing lines representing the 
original final-state system. 
\label{fig:labels}}
\end{figure}
Throughout this Letter, we use the following notation convention: 
 capital letters for pre-branching (parent) partons and
lower-case letters for post-branching (daughter) ones. Also, we use
the first letters of the alphabet, $a,b,c,...$, for incoming partons
and letters starting from $i,j,k,...$ for outgoing ones. 
Fig.~\ref{fig:labels} illustrates these choices for
the two basic types of configurations we consider.  
We will also mark incoming particles with a minus sign in front 
in antenna functions.
We adopt the
convention that particle energies are always positive, whether the 
particle is in the initial or the final state.  As a result, 
$s_{ij} = (k_i+k_j)^2$ is always positive.

The key building block for parton showers is the Sudakov factor, which
represents the non-emission probability between two values 
of the evolution scale, see
\cite{Buckley:2011ms,Beringer:1900zz} for reviews. In the context of
an antenna shower, the Sudakov factor for the branching of  one 
antenna is
\begin{equation}
	\Delta \left( Q_\text{start}^2, Q_\text{emit}^2 \right) 
= \exp \left[ - \antint ( Q_\text{start}^2, Q_\text{emit}^2 ) \right]\,,
\label{AntennaNonEmission}
\end{equation}
with
\begin{multline}
  \antint ( Q_\text{start}^2, Q_\text{emit}^2 ) = \\
  \int_{Q_\text{start}^2}^{Q_\text{emit}^2}
  \antc\, \frac{ f_a (x_a, Q^2 ) }{ f_A (x_A, Q^2 ) } 
     \frac{ f_b (x_b, Q^2 ) }{ f_B(x_B, Q^2) }
	\dd \Phi_\text{ant}\,.
\label{AntennaIntegral}
\end{multline}
In this equation, $\dd \Phi_\text{ant}$ represents the 
antenna phase-space factorization, which provides an exact Lorentz-invariant
mapping from 2 to 3 on-shell partons, that conserves 
global energy and momentum. Specific forms appropriate to
initial--final and initial--initial antenna configurations are defined
in sections \ref{sec:if} and \ref{sec:ii}, respectively. 

The evolution variable $Q^2$ is a function of the phase-space point
and must vanish in the unresolved limits~\cite{Skands:2009tb}.
The general formalism permits us to study different evolution 
variables~\cite{Giele:2007di,Giele:2011cb}, though in this Letter we
will restrict ourselves to a transverse-momentum type variable, 
defined in section \ref{sec:implementation}.   As in all parton
showers, the description is expected to be accurate only in the
strongly-ordered limit for the $Q^2$ of successive emissions.

The dressed or colored
antenna function $\antc$ is defined as\footnote{
Note that in \cite{Giele:2011cb} the normalization was
$\antc = \alphaS/(4 \pi) C \bar{a}$ 
}
\begin{equation}
	\antc = 4 \pi \alphaS ( Q^2) C \ordant\,,
\end{equation}
where $C$ is a color factor (we recall that we use normalization
conventions such 
that gluon and quark emission antenn\ae{} have $C=C_A$ and $C=2C_F$,
respectively, and gluon-splitting ones have $C=1$), and $\ordant$ is 
a color-ordered antenna function, which embodies the factorization of
QCD matrix elements in all single-unresolved soft and collinear limits. 
We don't take the functions $\ordant$ to be fixed;
instead we use different antenna functions with the same singular 
limits as one estimate of the shower uncertainty.

We use so-called global antenna
functions~\cite{Gustafson:1987rq}
(called sub-antenna functions with uniquely identified 
radiators in ref.~\cite{GehrmannDeRidder:2005cm}) 
which are active over all of 
phase space.
A backwards-evolution shower based on sector antenn\ae{} 
in analogy to refs.~\cite{Larkoski:2009ah, LopezVillarejo:2011ap} is left for
future work.
Some, but not all, antenn\ae{} needed for initial-state radiation
can be chosen to be the
crossings of their final--final counterparts.
An incoming particle is necessarily a hard radiator in an antenna.
Therefore, a gluon emission antenna function with an incoming gluon 
has to reproduce the AP splitting function on its own, e.\ g.\ 
\begin{equation}
	\ordant \left( -a_g, j_g, k_x \right)
		\xrightarrow{p_j \to z p_a}
		\frac{1}{s_{aj}} \frac{1}{1-z} P_{gg \to G}(1-z)
\end{equation}   
whereas if both gluons are in the final state, the collinear singularity 
is reproduced by the sum of two antenna functions
\begin{equation}
	\ordant \left( h_x, i_g, j_g \right) + \ordant \left( i_g, j_g, k_x \right)
	\xrightarrow{ p_j \to z p_i }
	\frac{1}{s_{ij}} P_{gg \to G}(z)
\end{equation}
 where the first antenna function is singular for $i$ becoming soft, 
 the second for $j$.

In pure final-state showers, the $x$ values of the incoming partons 
are not modified by the
phase-space factorization, hence the PDF ratios in eq.~\eqref{AntennaIntegral}
drop
out, yielding the ordinary form of the final--final Sudakov
form factor~\cite{Giele:2007di,Giele:2011cb}.  

For initial--final antenn\ae{}, only one of the PDF $x$ values changes,
and a Sudakov factor very similar to that of conventional AP showers
results, with a single PDF ratio in the kernel, $f_a(x_a,Q^2)/f_A(x_A,Q^2)$. 
Unlike conventional showers, however, we must also consider the
backwards evolution of two initial-state partons simultaneously,
generally requiring two separate parton-density factors in initial--initial
antenn\ae{}.

The consideration of initial--initial and initial--final antenn\ae{}
gives rise to one more subtlety.  The basic antenna functions
are color-ordered, so that in a final--final gluon-emission
antenna, for example, the emitted gluon is color adjacent to
both other (hard) daughter partons.  That is, it is the middle
parton of the color trio which is emitted.  The leading-color approximation
inherent in parton showers along with the symmetry of final-state
phase space allows us only antenn\ae{} with this ordering.  When
considering initial-state antenn\ae{}, however, the emitted parton need
not be color-adjacent to both other daughter partons; the middle parton,
adjacent to both, may end up in the initial instead of the final state.
We will call antenn\ae{} in which the middle parton is emitted into the final
state, `emission'
antenn\ae{}; and those in which the middle parton ends up in the initial
state, `conversion' antenn\ae{}.

For those antenn\ae{} in which the type (spin) of the initial-state partons
does not change after branching, we can redistribute collinear singularities
to neighboring antenn\ae{} so as to replace
`conversion' antenn\ae{} by `emission' antenn\ae{}.  For those
antenn\ae{} in which the type of the initial-state partons changes during
branching --- in which a quark backwards-evolves into a gluon or vice versa ---
we cannot avoid a consideration of both types of antenna function and
non-emission probability.

\section{Initial--Final Configurations \label{sec:if}}

The pre- and post-branching partons for initial--final configurations
are labeled by $AK \to ajk$, with the other incoming parton, $B$,
acting as a passive spectator, see the illustration in 
fig.~\ref{fig:labels}.   

In general, the incoming momentum after branching will no longer be
parallel to the beam direction.  We could boost it back to the beam
direction; this will transfer some of the transverse momentum generated
in the emission to the rest of the event.  This is the antenna analog
of the recoil considered in ref.~\cite{Platzer:2009jq}.
In the present Letter, we will instead restrict the branching so that
the incoming momentum remains parallel to the beam axis after branching.

With this restriction, the phase-space factorization reads~\cite{Catani:1996vz},
\begin{multline}
  \label{eq:factorizationif}
 	\int \frac{ \dd x_a}{ x_a} \dd \Phi_3 \left( -a,-c ; j, k, R \right) = 
	\\
	\int \frac{ \dd x_A }{ x_A } \dd \Phi_2 \left( -A, -c ; K, R \right) 
	\\
	\dd \Phi_\text{ant}^{if} \left(-A;K \to -a; j, k \right)
\end{multline}
with $x_A/x_a = \sAK/(\sAK+\sjk)$ and where 
the initial--final antenna phase space is
\begin{multline}
\dd \Phi_\text{ant}^{if} \left(-A;K \to -a; j, k \right) = 
	\\
	\frac{1}{ 16 \pi^2} \frac{ \sAK }{ (\sAK+\sjk)^2 } \dd \sjk \dd \saj 
\end{multline}
with the boundaries $0 \leq \sjk \leq \sAK (1-x_A)/x_A$, 
$0 \leq \saj \leq \sAK + \sjk$.
We have suppressed the integration over the third coordinate of the 
initial--final phase space on which the emission probability does not depend.

The gluon-emission antenn\ae{} can be chosen as
\begin{equation}
	\ordant ( -a_q, j_g, k_q ) = 
		\frac{1}{s_{AK}} \left( \frac{ 2 s_{ak} s_{AK} }{ s_{aj} s_{jk} } + \frac{s_{jk}}{s_{aj}} + \frac{s_{aj}}{s_{jk}}  \right)
\end{equation}

\begin{multline}
	\ordant ( -a_q, j_g, k_g ) = 
	\\
	\frac{1}{s_{AK}} \left( 
		\frac{ 2 s_{ak} s_{AK} }{ s_{aj} s_{jk} } + \frac{ s_{jk} }{ s_{aj}} + \frac{ s_{aj} }{ s_{jk}} \frac{ s_{ak} }{ s_{AK}}
	\right)
\end{multline}

\begin{multline}
	\ordant ( -a_g, j_g, k_g ) = 
		\frac{1}{ s_{AK} } \left(
			\frac{2 s_{ak} s_{AK} }{ s_{aj} s_{jk} } + \frac{ 2 s_{jk} s_{AK} }{ s_{aj} ( s_{ak} + s_{aj}) } 
			\right. \\ \left.
			+ \frac{ 2 s_{jk} }{ s_{aj} } \frac{ s_{ak} }{ s_{AK}} + \frac{ s_{aj} }{s_{jk} } \frac{ s_{ak} }{ s_{AK} }
		\right)
\end{multline}

\begin{multline}
	\ordant ( -a_g, j_g, b_q ) = \frac{1}{ s_{AK}} \left( 
			\frac{2 s_{ak} s_{AK} }{ s_{aj} s_{jk} } + \frac{ 2 s_{jk} s_{AK} }{ s_{aj} ( s_{ak} + s_{aj}) } 
			\right. \\ \left.
			+ \frac{2 s_{jk} }{ s_{aj} } \frac{ s_{ak} }{ s_{AK}} 
			+ \frac{ s_{aj} }{s_{jk} }
	\right)  
\end{multline}
where it is apparent that the antenn\ae{} with an incoming quark are crossings of their 
final-state counterparts whereas the ones with incoming gluons have additional terms 
compared to their final-state counterparts to ensure the collinear singularity $a \parallel j$
is  taken into account properly.

The antenna for the splitting of a final-state gluon into a quark-antiquark pair is 
chosen as
\begin{equation}
 	\ordant ( -a_x, j_q, k_{\bar{q}} ) = \frac{1}{2} \frac{1}{s_{jk}} \frac{ s_{aj}^2 + s_{ak}^2 }{ s_{AK}^2 }
 \end{equation}
where the factor $1/2$ originates from the fact that the gluon is part of two antenn\ae{}.

The antenna governing the backwards-evolution of a gluon into a quark is
 \begin{equation}
 	\ordant ( -a_q, j_q, k_x )  = \frac{1}{2} \frac{1}{s_{aj}} \frac{ s_{ak}^2 + s_{jk}^2 }{ s_{AK}^2 }
 \end{equation}

For the reverse process of a sea quark backwards-evolving into a gluon, we use
 \begin{equation}
 	\ordant_\text{conv} ( j_q, -a_g, k_x )= 
		\frac{1}{s_{AK}} \left( \frac{ - 2 s_{jk} (s_{AK} - s_{aj}) }{ s_{aj} (s_{aj} + s_{ak}) } + \frac{s_{ak}}{s_{aj}} \right)
 \end{equation}	
 with a color connection $j-a-k$ at variance with the other antenn\ae{}.

\section{Initial--Initial Configurations \label{sec:ii}}

For initial--initial antenn\ae{}, we label the pre- 
and post-branching partons by $AB \to ajb$, see
fig.~\ref{fig:labels}.   In the initial--initial case, we must
necessarily have transverse momentum generated, which must be absorbed
by the rest of the event.  There are two ways of proceeding.  One can
allow the incoming partons to be shifted away from the beam direction 
after branching, and then boost back to a frame in which they are again
parallel to the beam direction.  Alternatively, one can fix the incoming
partons to be parallel to the beam direction, and balance the new transverse
momentum by boosting the rest of the event appropriately.  In both cases,
there is a freedom in how the longitudinal part of the emission momentum
is absorbed into the initial state.  This corresponds to a freedom in
relating the post-branching momentum fractions $x_{a,b}$ to the pre-branching
momentum fractions $x_{A,B}$.  In the first case, this freedom is parametrized
by the recoil or reconstruction function $r$ in combination with the
Lorentz transformation boosting back to the lab frame.  
In the second case, it is parametrized by the functional form of $x_{a,b}$.

It turns out that these two approaches are equivalent, unlike the
initial--final case.  We define our recoil strategy in terms of $x_{a,b}$ here.  
The phase-space factorization reads~\cite{Catani:1996jh},
\begin{multline}
\int \frac{ \dd x_a }{x_a} \frac{ \dd x_b }{x_b} 
	\dd \Phi_2 \left( -a, -b ; j, R \right) 
	\\
	= \int \frac{ \dd x_A }{x_A} \frac{ \dd x_B }{x_B} 
	\dd \Phi_1 \left( -\lt{A}, -\lt{B} ; \lt{R} \right)
	\dd \Phi_\text{ant}^{ii}
\end{multline}
with the initial--initial antenna phase space
\begin{multline}
\dd \Phi_\text{ant}^{ii}\left(- \lt{A}, -\lt{B}   \to -a, -b ; j \right) = \\
		 \frac{1}{16 \pi^2} \frac{\sAB}{ \sab^2} 
		  \theta \left( 1- x_a \right) \theta\left( 1-x_b \right) \dd \saj \dd \sjb
\end{multline}
where we have suppressed the integration over the angle $\phi$ 
parametrizing rotations around the beam.
The pre- and post-branching momenta are related by a Lorentz 
transform:
\begin{multline}
	\Lambda\indices{^\mu_\nu} \left( p_R, \lt{p_R} \right)= 
	g\indices{^\mu_\nu} +\frac{2}{m_R^2} (\lt{p_R})^\mu (p_R)_\nu \\
	- \frac{2}{(p_R+\lt{p_R})^2} (p_R+\lt{p_R})^\mu (p_R+\lt{p_R})_\nu ~.
\end{multline}

The phase space boundary depends directly on the definition of the post-branching
momentum fractions, which is not fixed completely by the requirements of
$x_a x_b \sAB = x_A x_B \sab$ and the behavior in the soft and collinear limits.

For gluon emission, we use
\begin{equation}
	\frac{x_A}{x_a} = \left( \frac{ \sab - \sjb }{ \sab - \saj } \frac{ \sAB }{ \sab } \right)^{1/2}\mskip -15mu,
\end{equation}
whereas for conversion, we keep one incoming momentum fixed, \ie
$x_A/x_a = \sAB/\sab$, $x_b = x_B$, giving the phase space boundaries
$\saj + \sjb \leq \sAB (1-x_A)/x_A$.
This corresponds to the use of a one-sided factorization, which is possible
because only one collinear limit is singular, whereas the behavior of the 
phase space factorization in the other collinear limit is not constrained.

We use the emission antenn\ae{}
\begin{multline}
	\ordant \left( -a_q, j_g, -b_{\bar{q}} \right) = \frac{1}{s_{AB}} \left(
		 \frac{ 2 s_{ab} s_{AB} }{ s_{aj} s_{jb} } + \frac{ s_{aj} }{s_{jb}} + \frac{ s_{jb}}{s_{aj}} 
	\right)
\end{multline}
\begin{multline} 
	\ordant \left( -a_q, j_g, -b_g \right) = \frac{1}{s_{AB}} \left(
		 \frac{ 2 s_{ab} s_{AB} }{ s_{aj} s_{jb} } 
		 + \frac{ 2 s_{aj} s_{AB} }{ s_{jb} (s_{ab} + s_{jb}) }
		 \right. \\ \left.
		+  \frac{2 s_{aj} }{s_{jb}} \frac{ s_{ab}}{s_{AB}} + \frac{ s_{jb}}{s_{aj}} 
	\right)
\end{multline}
\begin{multline} 
	\ordant \left( -a_g, j_g, -b_g \right)  = \frac{1}{s_{AB}} \left(
		 \frac{ 2 s_{ab} s_{AB} }{ s_{aj} s_{jb} } 
		+ \frac{ 2 s_{aj} s_{AB} }{ s_{jb} (s_{ab} + s_{jb}) }
		\right. \\ \left.
		+ \frac{ 2 s_{jb} s_{AB} }{ s_{aj} (s_{ab} + s_{aj}) }
		+  \frac{2 s_{aj} }{s_{jb}} \frac{ s_{ab}}{s_{AB}} 
		+  \frac{2 s_{jb}}{s_{aj}} \frac{s_{ab}}{s_{AB}}
	\right)
\end{multline}

For a quark backwards-evolving into a gluon, we use
 \begin{equation}
 	\ordant \left( j_q, -a_g, -b_x \right) = \frac{1}{s_{AB}} 
	\left( - \frac{ 2 s_{jb} s_{AB} }{ s_{aj} (s_{ab} - s_{aj})} + \frac{s_{ab}}{s_{aj}} \right)
 \end{equation}

The antenna for a gluon backwards-evolving into a quark is 
the crossing of the initial--final counterpart:
\begin{equation}
	\ordant \left(-a_x, j_q, -b_q \right) = \frac{1}{2} \frac{1}{ s_{jb} } \frac{ s_{aj}^2 + s_{ab}^2 }{ s_{AB}^2 }
\end{equation}

\section{Implementation and Preliminary Results \label{sec:implementation}}

\def\trialant{\antc^{\rm trial}}
\def\Rpdf{R_{\rm pdf}}
\def\trialRpdf{\Rpdf^{\rm trial}}

In the antenna shower, as in a conventional shower, we start the
evolution at high $Q^2$, and generate a series of branchings at successively
lower $Q^2$, stopping when we reach a shower cut-off, typically 
around 1~GeV.  Each branching is generated according to the 
non-emission probability~(\ref{AntennaNonEmission}), and in this
Letter we shall restrict ourselves to strict strong ordering,
postponing a discussion of smooth ordering~\cite{Giele:2011cb} and/or power
showers~\cite{Plehn:2005cq} to a subsequent study.

In order to generate a branching, we must invert the function specified
by the integral~(\ref{AntennaIntegral}).  This is in general a difficult task
even if the integral is doable analytically, 
because the result involves dilogarithms.  In some cases, the boundaries
even make it unreasonable to perform the integral analytically.  A direct
inversion would in either case be quite slow.  Instead, we proceed as
follows.  We pick a simple function --- a trial antenna function 
$\trialant$ and trial ratios of parton-density functions $\trialRpdf$
---
which {\it overestimates\/} the
integrand, and veto the excess emissions generated according to the 
non-emission probability computed using the trial function.  The trial
function is chosen to capture the leading logarithmic singularities
of the antenna function, and to allow the phase-space integral
to be factorized into a product of one-dimensional integrals.  Where
possible, it is also chosen to produce an analytically invertible
integral.  In the final--final case, the latter requirement can always
be satisfied; in initial--initial and initial--final cases, it can be
satisfied for most trial antenn\ae{}.  In the exceptional cases, we
employ a two-stage veto.  In these cases,
 the first-level trial function still serves
to simplify the inversion of the non-emission probability by ensuring
the factorization of the integral into a product of one-dimensional integrals.
The veto probability is given by,
\begin{equation}
P_{\rm accept} = \frac{\antc}{\trialant} \frac{\Rpdf}{\trialRpdf}\,,
\end{equation}
evaluated at the post-branching kinematic point.  In this equation,
\begin{equation}
\Rpdf =  \frac{ f_a (x_a, Q^2 ) }{ f_A (x_A, Q^2 ) } 
     \frac{ f_b (x_b, Q^2 ) }{ f_B(x_B, Q^2) }
\end{equation}
is the ratio of parton densities that appears in eq.~\eqref{AntennaIntegral}.

When approaching a heavy flavor threshold, $\Rpdf$ diverges. 
As is standard in backwards-evolution codes, we absorb the leading divergent 
behavior of $\Rpdf$ into the trial integration to maintain a
reasonable efficiency~\cite{Sjostrand:2004ef}. 
Analogous issues may arise at large $x$ and low $Q^2$ in light-quark
parton densities due to numerical instabilities.
We defer their treatment to future work.  

To define a concrete shower algorithm based on the above antenn\ae{} and
phase-space factorizations, we have chosen to use two different
evolution variables, depending on the type of antenna.  For gluon
emission, we use a transverse momentum,
\begin{equation}
Q_\perp^2 = \frac{2 s_{ij} s_{jk}}{s_{ij} + s_{jk} + s_{ik}}~.
\end{equation}
For final-state branchings, $Q_\perp^2$ is equal to $2p^2_{\perp A}$,
the evolution measure used in \Ar{}~\cite{Lonnblad:1992tz} 
(note: previous \Vc{} publications used $4p_{\perp A}^2$). 
The maximal value of $Q_\perp^2$ in the final-state case is
$s_{ijk}/2$. 
For conversion and gluon splitting antenn\ae{}, we instead use the 
virtuality of the only potentially singular propagator as the 
evolution variable.
As in the final-state 
shower~\cite{Giele:2007di,GehrmannDeRidder:2011dm,Giele:2011cb},
other choices are possible within the \Vc{} formalism.  We defer
an exploration of more general possibilities to future work.

We now turn to a few basic tests of each component of the 
shower algorithm. In all cases, we consider $pp$ collisions at 8 TeV
CM energy, use the MSTW 2008 LO PDF set~\cite{Martin:2009iq}, with a one-loop
running $\alpha_S$, normalized to $\alpha_S(m_Z)=0.13939$. 
 In all calculations performed here, we turn off
hadronization and primordial $\kT$,
as well as the underlying event, both in the \Vc{} calculation
and in the \Pythia~8~\cite{Sjostrand:2007gs} 
calculations to which we compare.
While the evolution
variables in \Vc{} and \Pythia{} are different, we have tried to match
the shower cut-offs in calculations with the latter to that we use in
\Vc. This includes accounting for the difference in normalization
between limiting definitions of transverse momentum.
Note that while \Vc{} uses a zero-mass variable flavor number scheme, 
\Pythia~8 uses the physical quark 
masses everywhere. For the observables we discuss here, the effect is
negligible. 

\begin{figure}[t]
\centering
    \includegraphics[width = 0.49\textwidth]{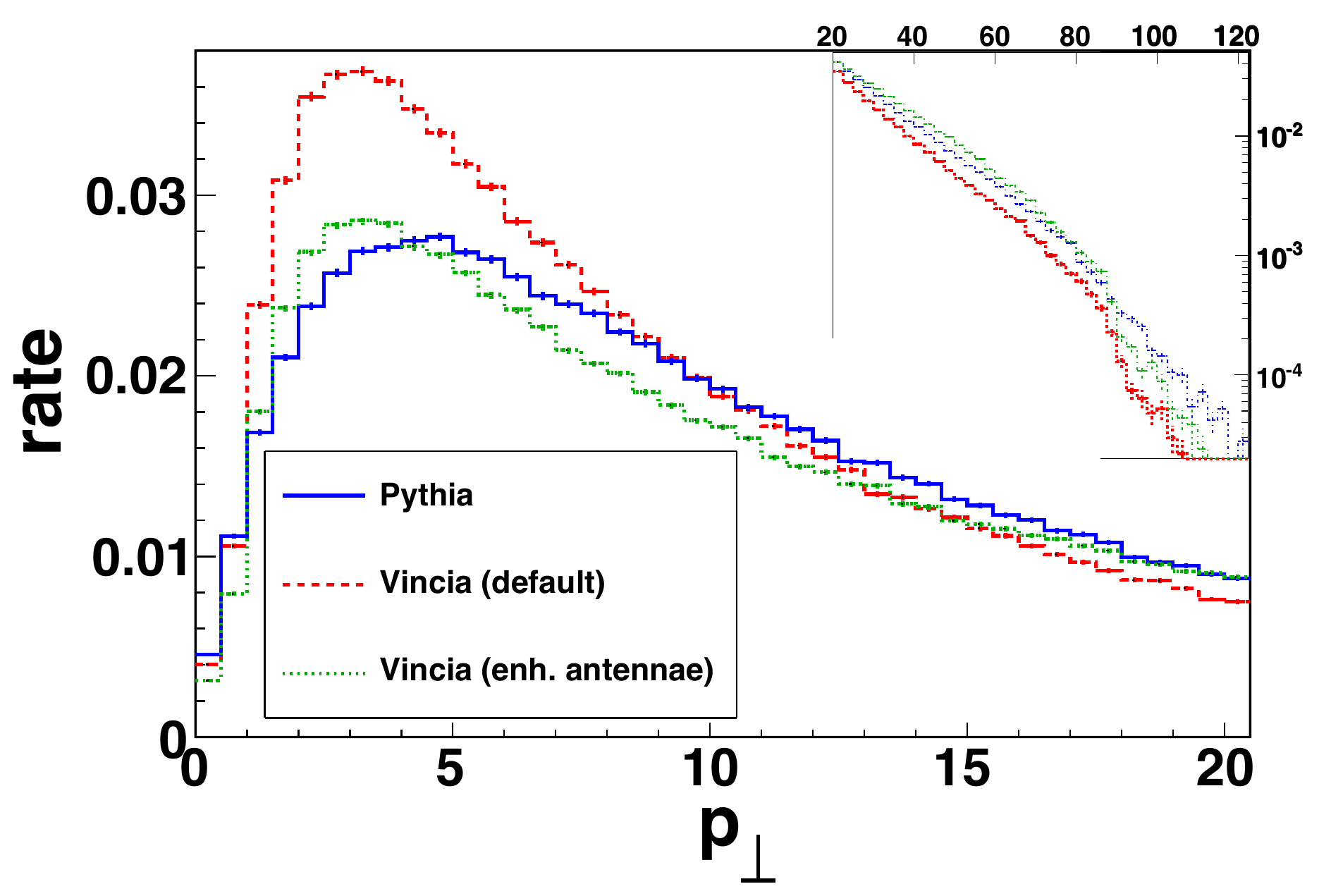}
\caption{The Drell-Yan $p_{T}$ spectrum.  The dashed red curve shows
the value computed using \Vc{} with default antenn\ae{} functions, while 
the dotted green curve shows the \Vc{} predicted with an enhanced
antenna function.  The solid blue curve gives the \Pythia~8 prediction.
The inset shows the high-$\pT$ tail.
\label{fig:DY}}
\end{figure}
In fig.~\ref{fig:DY}, we show the 
$p_T$ spectrum of the $Z$ boson in Drell-Yan production, which is
sensitive to radiation in initial--initial
configurations. 
The main figure pane
shows the peak of the distribution, while the inset shows
the high-$p_T$ tail.
The figure
shows \Vc{} curves computed using two different antenna functions:
the default antenna given earlier, and an enhanced antenna function,
with a finite term --- $5/s_{ijk}$ --- added.   
It also shows the result obtained with \Pythia~8. 

The overall shape of the three curves is similar: small values at
small $\pT$, rising to a peak and then declining again with a rough
power-law fall in the asymptotic region, and a ``knee'' around
$\pT\sim 90$ GeV due to the requirement of strong ordering which we
have imposed here. 
The difference between the
two \Vc{} predictions illustrates the uncertainty due to the shower
function and in particular higher-order terms in the shower.  The
difference shown here is illustrative only; a more extensive
exploration of possible antenna variations 
would be required before taking the spread as a quantitative estimate
of the uncertainty.  We may nonetheless observe that the \Pythia~8
reference calculation differs from the \Vc{} one (with default
antenna) by roughly the same amount in the peak region as does the
enhanced \Vc{} prediction. This illustrates a tradeoff between a more
active recoil strategy (\Pythia) and a more active radiation pattern
(enhanced \Vc{}), which will be interesting to study more closely. 
At large $p_T$, all three curves are close
to each other; the transverse momentum here is dominated by the recoil
against hard lone-gluon emission.  This region would be
described well by fixed-order calculations.  

\begin{figure}[t]
\centering
\includegraphics*[scale=0.2]{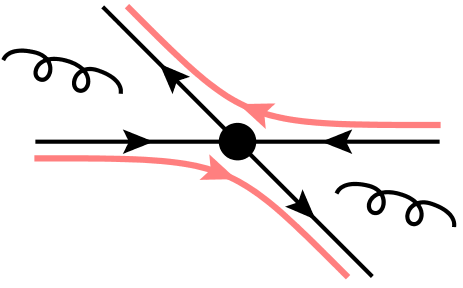} \hspace*{1mm}
\includegraphics*[scale=0.2]{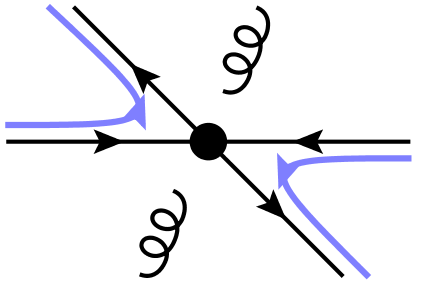}
\caption{Different color flows and corresponding emission patterns 
in $qq\to qq$ scattering.  The straight (black) lines are quarks with
arrows denoting the direction of motion in the initial or final states,
and the curved (colored) lines indicating the color flow.  The beam axis
is horizontal, and the vertical axis is transverse to the beam.  The 
initial-state momenta would be reversed in a Feynman diagram, so that
the gluon emissions symbolically indicated by curly lines would be inside
the corresponding color antenn\ae{}.  Forward flow is shown on the left,
and backward flow on the right.\label{fig:qq}}
\end{figure}

For initial--final configurations, coherence is particularly
important, and can lead to sizable
asymmetries (see, e.g.,~\cite{Skands:2012mm}). 
An illustration of the effect is given in fig.~\ref{fig:qq}, which
shows $qq \to qq$ scattering with two different color-flow assignments: 
forward (left) and backward (right). 
In both cases, the starting scale of the shower evolution would 
be $\hat{p}_T$, the transverse-momentum scale characterizing the hard
scattering. Coherence, however, implies 
that radiation should be directed primarily inside the color antenna,
so that in the forward flow it would be directed towards large rapidity,
and strongly suppressed 
at right angles to the beam direction.
In the backward flow, conversely, radiation at right angles to the
beam should be unsuppressed. 
The two radiation patterns are illustrated schematically 
by the gluons in fig.~\ref{fig:qq}. 
\begin{figure}[t]
\centering
\includegraphics*[scale=0.43]{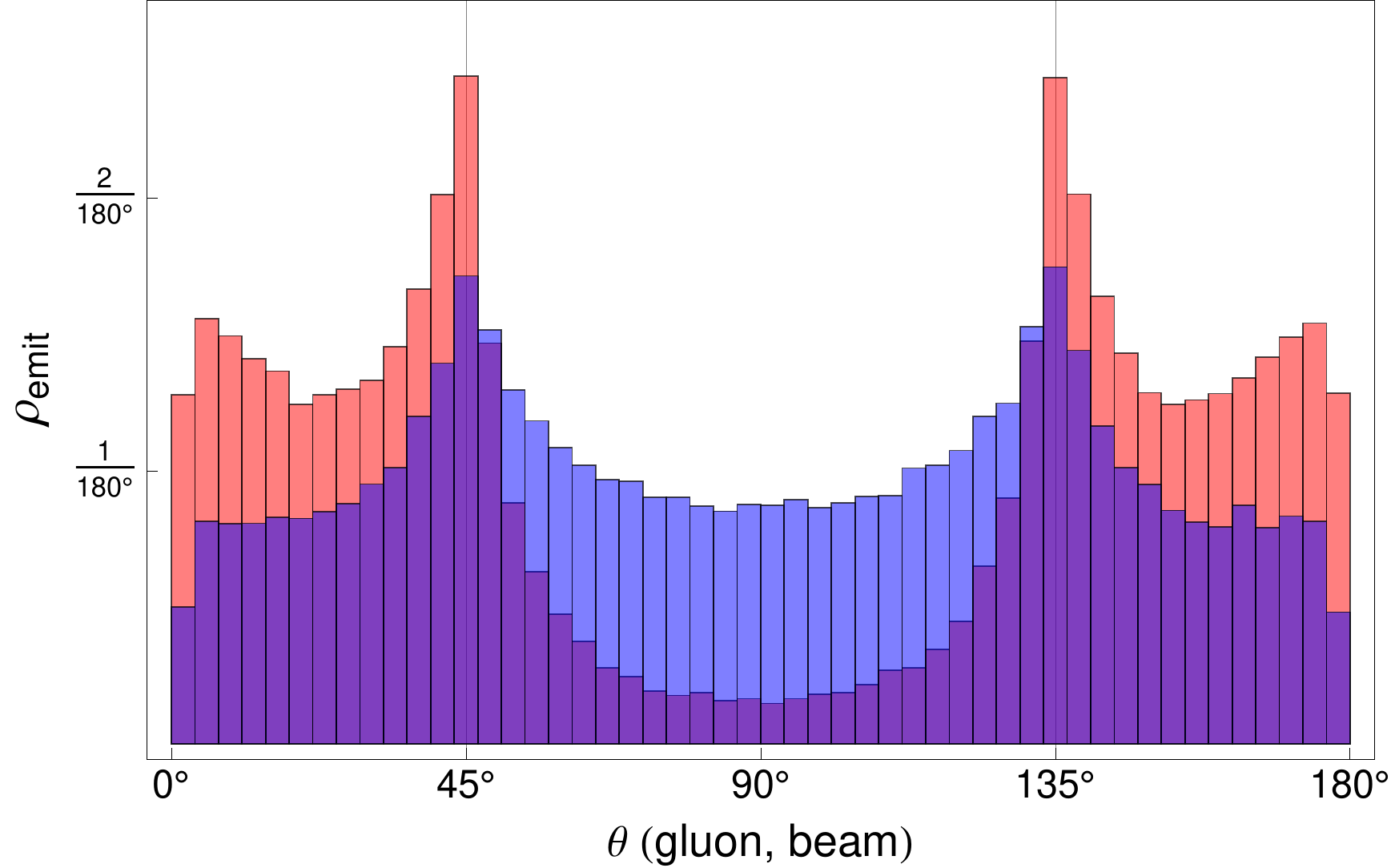}
\caption{Angular distribution of the first gluon emission in $qq\to
  qq$ scattering at $45^\circ$, for the two different color flows. 
The light (red) histogram shows the emission density for the forward
flow, and the dark (blue) histogram shows the emission density for the
backward flow.
\label{fig:coherence}}
\end{figure}
The intrinsic coherence of the antenna formalism accounts for this
effect automatically. That \Vc{} reproduces this feature 
is demonstrated in fig.~\ref{fig:coherence},
which shows the angular distribution of the first emitted gluon for
the forward and backward color flows, respectively, for a scattering
angle of 45$^\circ$ and $\hat{p}_{\rm T} = 100\,$GeV.  The
distributions clearly show that the backward color flow allows for
much more radiation at $90^\circ$ than the forward one. The $\pT$
spectrum of the radiation (not shown) is also harder. The next step
will be to interface the hadronization and underlying-event
models in \Pythia, and compare to experimental studies, 
such as the one by CDF~\cite{Abe:1994nj} (we note that an update of 
that study, correcting it to the hadron level, would be highly useful
to the MC community).

\begin{figure}[t]
\centering
\includegraphics[width=0.49\textwidth]{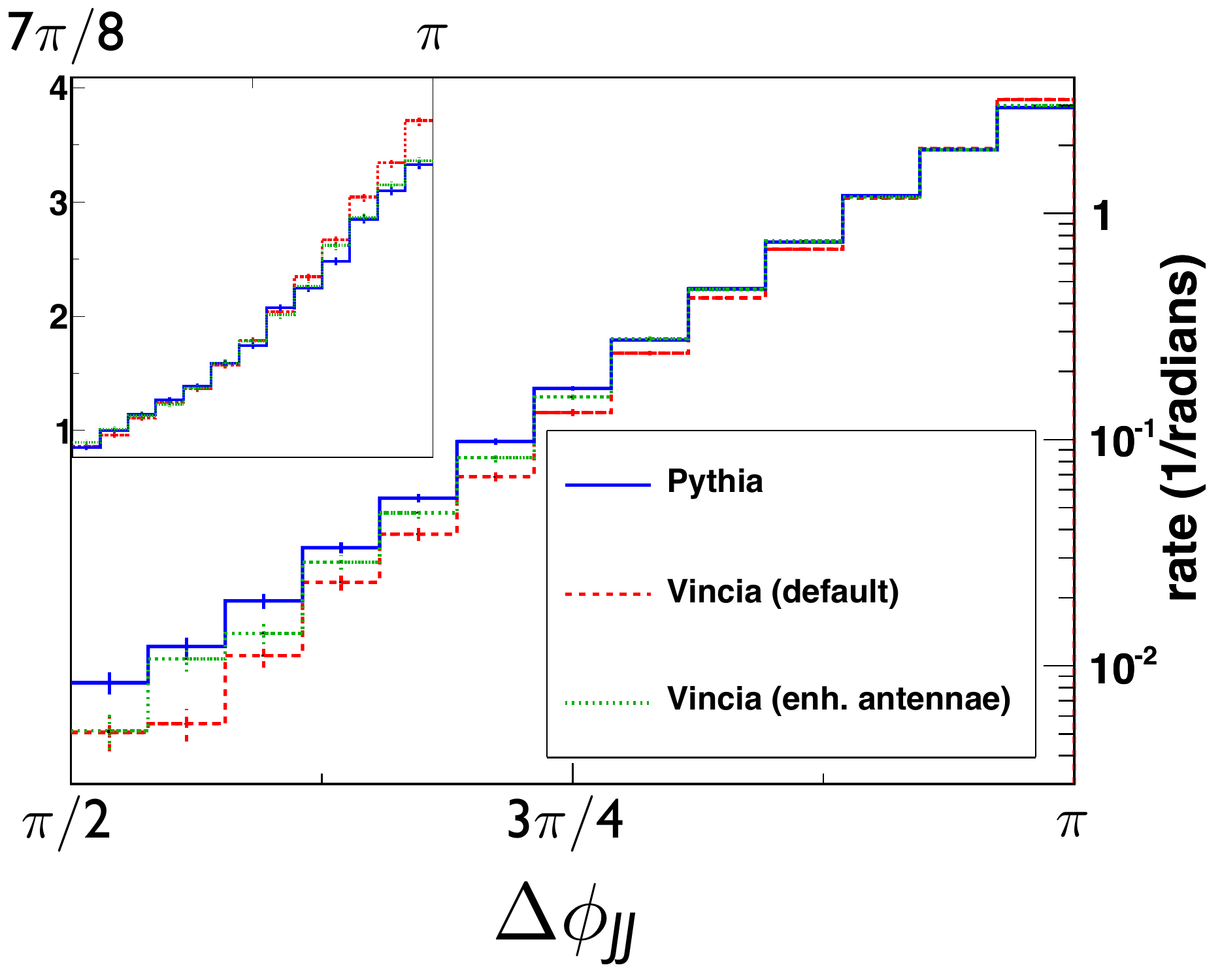}
\caption{The dijet decorrelation angle. \label{fig:dijet}
The histograms are normalized to unity separately.
}

\end{figure}
Finally, to demonstrate the combination of all shower components
acting together, we show the dijet decorrelation angle (the azimuthal
angle between the two leading jets), $\Delta\phi_{JJ}$, in
fig.~\ref{fig:dijet}.  Using \Fastjet~\cite{Cacciari:2011ma}, we
consider anti-$\kT$ jets with radius parameter $R=0.4$. We demand the
two leading jets to have transverse momentum above $100$ GeV and to be
at rapidities $\abs{y} < 2.8$.  Note that \Pythia~8 produces about $40
\% $ more jet events which pass these cuts than \Vc{} both with or
without enhanced antenna functions,
partly due to its more active recoil strategy, 
which allows the original dijet system 
to build up transverse momentum successively during the shower cascade.
This difference is not visible in fig.~\ref{fig:dijet}, as the
distributions are all normalized to unity.  The two \Vc{}
distributions are broadly similar to the \Pythia~8 distribution, and
very similar in the two-jet region ($\Delta\phi_{JJ}\sim\pi$) where
the parton-shower approximation should be reliable.  The differences
are substantial in the region below $\Delta\phi_{JJ}<3\pi/4$, where
hard real emission is important.  In this region, fixed-order
calculations should be reliable, but unmatched parton showers will not
be.  Nonetheless, the difference between the \Vc{} calculation with
default antenna strength and the \Pythia~8 calculation is similar to
that between the two \Vc{} calculations, suggesting again that the
variation provides a good qualitative assessment of the uncertainty.
We defer a comparison of this distribution with fixed-order
calculations to future studies.

\section{Conclusion and Outlook \label{sec:conclusion}}

In this Letter, we have presented the outline of a formalism for an antenna
shower for hadron collisions, along with results from an initial
implementation as a plug-in to \Pythia~8.  The formalism requires the
introduction of new antenn\ae{}, corresponding to one or both parents being 
initial-state partons.  These should be further subdivided into the new
categories of emission and conversion antenn\ae{} based on their color flows. 
The formalism also requires factorizations suitable for phase spaces
involving initial-state partons, and introduces ratios of parton densities
into the non-emission probabilities governing the shower evolution. 

We have chosen to implement the shower as a plug-in to the \Pythia~8 
program, which takes advantage of the latter's flexible framework and
makes use of its utilities, structures, and overall management of the
branching process.  In this approach, it replaces \Pythia's shower with an
antenna shower. For practical and efficiency reasons, we uniformly adopt
a trial-and-veto algorithm for generating branchings.  The trial functions
used in the implementation will be described elsewhere.

We expect to implement further optimizations of the branching step 
in future work.  The leading-order matching approach described in
ref.~\cite{Giele:2011cb} should carry over to the initial-state showering
described here, and will be an important next step for the development
of \Vc.

\section*{Acknowledgments}
The work of MR and DAK was supported by
the European Research Council under Advanced Investigator Grant
ERC--AdG--228301.




\bibliographystyle{model1-num-names}
\bibliography{main}







\end{document}

%% file: notation.tex
\newcommand{\bit}{\begin{itemize}}
\newcommand{\eit}{\end{itemize}}
\newcommand{\bc}{\begin{center}}
\newcommand{\ec}{\end{center}}
\newcommand{\bq}{\begin{quote}}
\newcommand{\eq}{\end{quote}}
\newcommand{\bi}{\begin{itemize}}
\newcommand{\ei}{\end{itemize}}
\newcommand{\be}{\begin{equation}}
\newcommand{\ee}{\end{equation}}
\newcommand{\bea}{\begin{eqnarray}}
\newcommand{\eea}{\end{eqnarray}}
\newcommand{\bt}{\begin{tabular}}
\newcommand{\et}{\end{tabular}}
\newcommand{\bmp}{\begin{minipage}}
\newcommand{\emp}{\end{minipage}}
\newcommand{\btab}{\begin{tabbing}}
\newcommand{\etab}{\end{tabbing}}

\newcommand{\beq}{\begin{equation}}
\newcommand{\eeq}{\end{equation}}
\newcommand{\beqa}{\begin{eqnarray}}
\newcommand{\eeqa}{\end{eqnarray}}
\def\trm{\textrm}

\def\ie{\trm{i.e.\ }}

\def\eq#1{eq.~(\ref{#1})}

\newcommand{\abs}[1]{\lvert#1\rvert}

\newcommand{\dd}{\text{d}}

\newcommand{\antint}{\ensuremath{\mathcal{A}}}
\newcommand{\alphaS}{\ensuremath{\alpha_S}}

\newcommand{\pT}{\ensuremath{p_\perp}\xspace}

\newcommand{\sAK}{\ensuremath{s_{AK}}\xspace}
\newcommand{\saj}{\ensuremath{s_{aj}}\xspace}
\newcommand{\sjk}{\ensuremath{s_{jk}}\xspace}
\newcommand{\sAB}{\ensuremath{s_{AB}}\xspace}
\newcommand{\sjb}{\ensuremath{s_{jb}}\xspace}

\newcommand{\sab}{\ensuremath{s_{ab}}\xspace}

\newcommand{\lt}[1]{\ensuremath{#1}'}

\newcommand{\start}{\ensuremath{\text{start}}}
\newcommand{\emit}{\ensuremath{\text{emit}}}




%% file: main.bbl
\begin{thebibliography}{28}
\expandafter\ifx\csname natexlab\endcsname\relax\def\natexlab#1{#1}\fi
\providecommand{\bibinfo}[2]{#2}
\ifx\xfnm\relax \def\xfnm[#1]{\unskip,\space#1}\fi
\bibitem[{Buckley et~al.(2011)Buckley, Butterworth, Gieseke, Grellscheid,
  H{\"o}che et~al.}]{Buckley:2011ms}
\bibinfo{author}{A.~Buckley}, \bibinfo{author}{J.~Butterworth},
  \bibinfo{author}{S.~Gieseke}, \bibinfo{author}{D.~Grellscheid},
  \bibinfo{author}{S.~H{\"o}che}, et~al.,
\newblock \bibinfo{title}{{General-purpose event generators for LHC physics}},
\newblock \bibinfo{journal}{Phys.Rept.} \bibinfo{volume}{504}
  (\bibinfo{year}{2011}) \bibinfo{pages}{145}.
\bibitem[{Sj{\"o}strand(1985)}]{Sjostrand:1985xi}
\bibinfo{author}{T.~Sj{\"o}strand},
\newblock \bibinfo{title}{{A Model for Initial State Parton Showers}},
\newblock \bibinfo{journal}{Phys.Lett.} \bibinfo{volume}{B157}
  (\bibinfo{year}{1985}) \bibinfo{pages}{321}.
\bibitem[{Altarelli and Parisi(1977)}]{Altarelli:1977zs}
\bibinfo{author}{G.~Altarelli}, \bibinfo{author}{G.~Parisi},
\newblock \bibinfo{title}{{Asymptotic Freedom in Parton Language}},
\newblock \bibinfo{journal}{Nucl.Phys.} \bibinfo{volume}{B126}
  (\bibinfo{year}{1977}) \bibinfo{pages}{298}.
\bibitem[{Gustafson and Pettersson(1988)}]{Gustafson:1987rq}
\bibinfo{author}{G.~Gustafson}, \bibinfo{author}{U.~Pettersson},
\newblock \bibinfo{title}{{Dipole Formulation of QCD Cascades}},
\newblock \bibinfo{journal}{Nucl.Phys.} \bibinfo{volume}{B306}
  (\bibinfo{year}{1988}) \bibinfo{pages}{746}.
\bibitem[{L{\"o}nnblad(1992)}]{Lonnblad:1992tz}
\bibinfo{author}{L.~L{\"o}nnblad},
\newblock \bibinfo{title}{{ARIADNE version 4: A Program for simulation of QCD
  cascades implementing the color dipole model}},
\newblock \bibinfo{journal}{Comput.Phys.Commun.} \bibinfo{volume}{71}
  (\bibinfo{year}{1992}) \bibinfo{pages}{15}.
\bibitem[{Winter and Krauss(2008)}]{Winter:2007ye}
\bibinfo{author}{J.-C. Winter}, \bibinfo{author}{F.~Krauss},
\newblock \bibinfo{title}{{Initial-state showering based on colour dipoles
  connected to incoming parton lines}},
\newblock \bibinfo{journal}{JHEP} \bibinfo{volume}{0807} (\bibinfo{year}{2008})
  \bibinfo{pages}{040}.
\bibitem[{Kosower(1998)}]{Kosower:1997zr}
\bibinfo{author}{D.~A. Kosower},
\newblock \bibinfo{title}{{Antenna factorization of gauge theory amplitudes}},
\newblock \bibinfo{journal}{Phys.Rev.} \bibinfo{volume}{D57}
  (\bibinfo{year}{1998}) \bibinfo{pages}{5410}.
\bibitem[{Kosower(2005)}]{Kosower:2003bh}
\bibinfo{author}{D.~A. Kosower},
\newblock \bibinfo{title}{{Antenna factorization in strongly ordered limits}},
\newblock \bibinfo{journal}{Phys.Rev.} \bibinfo{volume}{D71}
  (\bibinfo{year}{2005}) \bibinfo{pages}{045016}.
\bibitem[{Gehrmann-De~Ridder et~al.(2005)Gehrmann-De~Ridder, Gehrmann, and
  Glover}]{GehrmannDeRidder:2005cm}
\bibinfo{author}{A.~Gehrmann-De~Ridder}, \bibinfo{author}{T.~Gehrmann},
  \bibinfo{author}{E.~N. Glover},
\newblock \bibinfo{title}{{Antenna subtraction at NNLO}},
\newblock \bibinfo{journal}{JHEP} \bibinfo{volume}{0509} (\bibinfo{year}{2005})
  \bibinfo{pages}{056}.
\bibitem[{Daleo et~al.(2007)Daleo, Gehrmann, and Maitre}]{Daleo:2006xa}
\bibinfo{author}{A.~Daleo}, \bibinfo{author}{T.~Gehrmann},
  \bibinfo{author}{D.~Maitre},
\newblock \bibinfo{title}{{Antenna subtraction with hadronic initial states}},
\newblock \bibinfo{journal}{JHEP} \bibinfo{volume}{0704} (\bibinfo{year}{2007})
  \bibinfo{pages}{016}.
\bibitem[{Giele et~al.(2008)Giele, Kosower, and Skands}]{Giele:2007di}
\bibinfo{author}{W.~T. Giele}, \bibinfo{author}{D.~A. Kosower},
  \bibinfo{author}{P.~Z. Skands},
\newblock \bibinfo{title}{{A Simple shower and matching algorithm}},
\newblock \bibinfo{journal}{Phys.Rev.} \bibinfo{volume}{D78}
  (\bibinfo{year}{2008}) \bibinfo{pages}{014026}.
\bibitem[{Gehrmann-De~Ridder et~al.(2012)Gehrmann-De~Ridder, Ritzmann, and
  Skands}]{GehrmannDeRidder:2011dm}
\bibinfo{author}{A.~Gehrmann-De~Ridder}, \bibinfo{author}{M.~Ritzmann},
  \bibinfo{author}{P.~Skands},
\newblock \bibinfo{title}{{Timelike Dipole-Antenna Showers with Massive
  Fermions}},
\newblock \bibinfo{journal}{Phys.Rev.} \bibinfo{volume}{D85}
  (\bibinfo{year}{2012}) \bibinfo{pages}{014013}.
\bibitem[{Lopez-Villarejo and Skands(2011)}]{LopezVillarejo:2011ap}
\bibinfo{author}{J.~Lopez-Villarejo}, \bibinfo{author}{P.~Skands},
\newblock \bibinfo{title}{{Efficient Matrix-Element Matching with Sector
  Showers}},
\newblock \bibinfo{journal}{JHEP} \bibinfo{volume}{1111} (\bibinfo{year}{2011})
  \bibinfo{pages}{150}.
\bibitem[{Bern et~al.(2008)}]{Bern:2008ef}
\bibinfo{author}{Z.~Bern}, et~al.,
\newblock \bibinfo{title}{{The NLO multileg working group: Summary report}}
  (\bibinfo{year}{2008}). \bibinfo{note}{ArXiv:0803.0494}.
\bibitem[{Giele et~al.(2011)Giele, Kosower, and Skands}]{Giele:2011cb}
\bibinfo{author}{W.~Giele}, \bibinfo{author}{D.~Kosower},
  \bibinfo{author}{P.~Skands},
\newblock \bibinfo{title}{{Higher-Order Corrections to Timelike Jets}},
\newblock \bibinfo{journal}{Phys.Rev.} \bibinfo{volume}{D84}
  (\bibinfo{year}{2011}) \bibinfo{pages}{054003}.
\bibitem[{Beringer et~al.(2012)}]{Beringer:1900zz}
\bibinfo{author}{J.~Beringer}, et~al.,
\newblock \bibinfo{title}{{Review of Particle Physics (RPP)}},
\newblock \bibinfo{journal}{Phys.Rev.} \bibinfo{volume}{D86}
  (\bibinfo{year}{2012}) \bibinfo{pages}{010001}.
\bibitem[{Skands and Weinzierl(2009)}]{Skands:2009tb}
\bibinfo{author}{P.~Z. Skands}, \bibinfo{author}{S.~Weinzierl},
\newblock \bibinfo{title}{{Some remarks on dipole showers and the DGLAP
  equation}},
\newblock \bibinfo{journal}{Phys.Rev.} \bibinfo{volume}{D79}
  (\bibinfo{year}{2009}) \bibinfo{pages}{074021}.
\bibitem[{Larkoski and Peskin(2010)}]{Larkoski:2009ah}
\bibinfo{author}{A.~J. Larkoski}, \bibinfo{author}{M.~E. Peskin},
\newblock \bibinfo{title}{{Spin-Dependent Antenna Splitting Functions}},
\newblock \bibinfo{journal}{Phys.Rev.} \bibinfo{volume}{D81}
  (\bibinfo{year}{2010}) \bibinfo{pages}{054010}.
\bibitem[{Pl{\"a}tzer and Gieseke(2011)}]{Platzer:2009jq}
\bibinfo{author}{S.~Pl{\"a}tzer}, \bibinfo{author}{S.~Gieseke},
\newblock \bibinfo{title}{{Coherent Parton Showers with Local Recoils}},
\newblock \bibinfo{journal}{JHEP} \bibinfo{volume}{1101} (\bibinfo{year}{2011})
  \bibinfo{pages}{024}.
\bibitem[{Catani and Seymour(1997)}]{Catani:1996vz}
\bibinfo{author}{S.~Catani}, \bibinfo{author}{M.~Seymour},
\newblock \bibinfo{title}{{A General algorithm for calculating jet
  cross-sections in NLO QCD}},
\newblock \bibinfo{journal}{Nucl.Phys.} \bibinfo{volume}{B485}
  (\bibinfo{year}{1997}) \bibinfo{pages}{291--419}.
\bibitem[{Catani and Seymour(1996)}]{Catani:1996jh}
\bibinfo{author}{S.~Catani}, \bibinfo{author}{M.~Seymour},
\newblock \bibinfo{title}{{The Dipole formalism for the calculation of QCD jet
  cross-sections at next-to-leading order}},
\newblock \bibinfo{journal}{Phys.Lett.} \bibinfo{volume}{B378}
  (\bibinfo{year}{1996}) \bibinfo{pages}{287--301}.
\bibitem[{Plehn et~al.(2007)Plehn, Rainwater, and Skands}]{Plehn:2005cq}
\bibinfo{author}{T.~Plehn}, \bibinfo{author}{D.~Rainwater},
  \bibinfo{author}{P.~Z. Skands},
\newblock \bibinfo{title}{{Squark and gluino production with jets}},
\newblock \bibinfo{journal}{Phys.Lett.} \bibinfo{volume}{B645}
  (\bibinfo{year}{2007}) \bibinfo{pages}{217--221}.
\bibitem[{Sj{\"o}strand and Skands(2005)}]{Sjostrand:2004ef}
\bibinfo{author}{T.~Sj{\"o}strand}, \bibinfo{author}{P.~Z. Skands},
\newblock \bibinfo{title}{{Transverse-momentum-ordered showers and interleaved
  multiple interactions}},
\newblock \bibinfo{journal}{Eur.Phys.J.} \bibinfo{volume}{C39}
  (\bibinfo{year}{2005}) \bibinfo{pages}{129--154}.
\bibitem[{Martin et~al.(2009)Martin, Stirling, Thorne, and
  Watt}]{Martin:2009iq}
\bibinfo{author}{A.~Martin}, \bibinfo{author}{W.~Stirling},
  \bibinfo{author}{R.~Thorne}, \bibinfo{author}{G.~Watt},
\newblock \bibinfo{title}{{Parton distributions for the LHC}},
\newblock \bibinfo{journal}{Eur.Phys.J.} \bibinfo{volume}{C63}
  (\bibinfo{year}{2009}) \bibinfo{pages}{189--285}.
\bibitem[{Sj{\"o}strand et~al.(2008)Sj{\"o}strand, Mrenna, and
  Skands}]{Sjostrand:2007gs}
\bibinfo{author}{T.~Sj{\"o}strand}, \bibinfo{author}{S.~Mrenna},
  \bibinfo{author}{P.~Z. Skands},
\newblock \bibinfo{title}{{A Brief Introduction to PYTHIA 8.1}},
\newblock \bibinfo{journal}{Comput.Phys.Commun.} \bibinfo{volume}{178}
  (\bibinfo{year}{2008}) \bibinfo{pages}{852}.
\bibitem[{Skands et~al.(2012)Skands, Webber, and Winter}]{Skands:2012mm}
\bibinfo{author}{P.~Skands}, \bibinfo{author}{B.~Webber},
  \bibinfo{author}{J.~Winter},
\newblock \bibinfo{title}{{QCD Coherence and the Top Quark Asymmetry}},
\newblock \bibinfo{journal}{JHEP} \bibinfo{volume}{1207} (\bibinfo{year}{2012})
  \bibinfo{pages}{151}.
\bibitem[{Abe et~al.(1994)}]{Abe:1994nj}
\bibinfo{author}{F.~Abe}, et~al.,
\newblock \bibinfo{title}{{Evidence for color coherence in $p\bar{p}$
  collisions at $\sqrt{s} = 1.8$ TeV}},
\newblock \bibinfo{journal}{Phys.Rev.} \bibinfo{volume}{D50}
  (\bibinfo{year}{1994}) \bibinfo{pages}{5562--5579}.
\bibitem[{Cacciari et~al.(2012)Cacciari, Salam, and Soyez}]{Cacciari:2011ma}
\bibinfo{author}{M.~Cacciari}, \bibinfo{author}{G.~P. Salam},
  \bibinfo{author}{G.~Soyez},
\newblock \bibinfo{title}{{FastJet User Manual}},
\newblock \bibinfo{journal}{Eur.Phys.J.} \bibinfo{volume}{C72}
  (\bibinfo{year}{2012}) \bibinfo{pages}{1896}.

\end{thebibliography}
